\def\changesHilighted{false} 
\documentclass{ifacconf}

\usepackage{graphicx}      
\usepackage{natbib}        
\usepackage[dvipsnames]{xcolor}
\usepackage{amsmath}
\usepackage{mathtools}
\usepackage{subfig}
\usepackage[normalem]{ulem}
\definecolor{americanrose}{rgb}{1.0, 0.01, 0.24}

\newcommand{\norm}[1]{\lVert #1 \rVert}

\newcommand{\anna}[1]{\textcolor{MidnightBlue}{\ [藤岡：#1]}}
\newcommand{\masaki}[1]{\textcolor{americanrose}{\ [小蔵：#1]}}

\newcommand{\del}[1]{\textcolor{red}{\sout{#1}}}
\newcommand{\addd}[1]{\textcolor{blue}{#1}}
\newcommand{\dell}[1]{\textcolor{red}{\sout{#1}}}

\ifnum\pdfstrcmp{\changesHilighted}{false}=0
    \renewcommand{\anna}[1]{}
    \renewcommand{\masaki}[1]{}
    \renewcommand{\del}[1]{}
    
    \renewcommand{\dell}[1]{}
    \renewcommand{\addd}[1]{#1}
\fi

\begin{document}
\begin{frontmatter}

\title{Shepherding Heterogeneous Flocks: Overview and Prospect\thanksref{footnoteinfo}} 

\thanks[footnoteinfo]{This work was supported by JSPS KAKENHI Grant Numbers JP21H01352, JP21H01353, JP22H00514 and JST Moonshot R\&D Grant Number JPMJMS2284.}

\author{Anna Fujioka${}^\ast$} \author{Masaki Ogura${}^{\ast\ast}$} 
\author{Naoki Wakamiya${}^{\ast\ast\ast}$}

\address{Department of Bioinformatic Engineering, Graduate School of Information Science and Technology, Osaka University, Osaka, Japan
email: ${}^\ast$a-fujioka@ist.osaka-u.ac.jp, 
${}^{\ast\ast}$m-ogura@ist.osaka-u.ac.jp, 
\mbox{${}^{\ast\ast\ast}$wakamiya@ist.osaka-u.ac.jp}}

\begin{abstract}                
The problem of guiding a flock of several autonomous agents using repulsion force exerted by a smaller number of agents is called the shepherding problem and has been attracting attention due to its potential engineering applications. Although several works propose methodologies for achieving the shepherding task in this context, most assume that sheep agents have the same dynamics, which only sometimes holds in reality. The objective of this discussion paper is to overview a recent research trend addressing the gap mentioned above between the commonly placed uniformity assumption and the reality. Specifically, we first introduce recent guidance methods for heterogeneous flocks and then describe the prospects of the shepherding problem for heterogeneous flocks. 
\end{abstract}


\end{frontmatter}

\section{Introduction}\label{sc:intro}
Flocking systems inspired by flocks of fish, birds, and other living animals are attracting attentions due to their potential applications in various areas of control engineering~\citep{hemelrijk2008self, ballerini2008empirical, mikami2020agent}
. Among these navigation, a navigation inspired by shepherds' behavior of guiding a flock of sheep is called shepherding~\citep{long2020comprehensive}. In shepherding, a small number of the shepherd agents guide a flock of sheep agents using their repulsion force to the sheep agents.
\addd{This guidance has many potential applications including crowd control~\citep{lien2009interactive} and oil spill cleanups~\citep{masehian2015cooperative}.}
The mathematical model of such guidance is called the shepherding model and the problem of guiding a flock of sheep agents is called the shepherding problem. The objective of this problem is to design a movement law of the shepherd agent for guiding the sheep agents. 

Although there exist several movement laws~\citep{tsunoda2019analysis, Deng2022}
, most of these studies assume that all sheep agents have the same dynamics. However, this assumption does not necessarily hold in several practical scenarios \citep{hemelrijk2008self, ballerini2008empirical, jolles2020role}. In this context, we can find in the literature a recent research trend in which the heterogeneity of the sheep agents is explicitly considered for designing the shepherding algorithm. 

Recently, \cite{himo2022iterative} proposed a guidance method for a flock of agents containing those which do not receive repulsion force from the shepherd. In this method, the shepherd guides the responsive sheep around an unresponsive sheep, and then, guides all sheep to a goal area using the sheep's behavior of making a flock. Furthermore, we recently expanded this method and proposed the guidance method for a flock containing variant sheep. In this method, the shepherd determine the type of the sheep based on the estimated dynamics. Then, the shepherd guide the sheep which are determined to be normal to the goal area. The objective of this paper is to describe these previous research works by the authors and, furthermore, provide future prospects on the problem of shepherding heterogeneous flocks. 

\section{PREVIOUS RESEARCHES}\label{sc:methods}

In this section, we describe the previous research works for shepherding a heterogeneous flock. First, in Subsection~\ref{subsc:shepherding_model}, we introduce a mathematical model of the dynamics of sheep agents. We then in Subsection~\ref{subsc:FAT} introduce a guidance method for shepherding a homogeneous flock, as this method forms a basis of the ones for a heterogeneous flock. We then describe two existing methodologies for shepherding heterogeneous flocks in Subsections~\ref{subsc:method_himo} and~\ref{subsc:method_fuji}.

\subsection{Shepherding Model}\label{subsc:shepherding_model}
First, we introduce the sheep agent model commonly used in the literature. The model is based on the Boid model proposed by~\cite{reynolds1987flocks}. Within the model, the position~$x_i(k)$ of the $i$th sheep agent at time~$k$ is updated by the difference equation 
\begin{equation}
    x_i(k+1) = x_i(k) + v_i(k),
\end{equation}
where $v_i(k)$ denotes the movement vector of the $i$th sheep at time~$k$. 
In the Boid model, sheep agents are assumed to receive the following three types of forces: ``separation force'' to prevent sheep from colliding with each other, ``alignment force'' to match the velocity of other agents, and ``attraction force'' to approach other agents. In addition to these forces, it is assumed that the sheep avoid the shepherd agent. Therefore, the movement vector of $i$th sheep at time~$k$ is given by
\begin{equation}\label{eq:sheep_vector}
\begin{multlined}[.8\linewidth]
    v_i(k) 
    = K_{i1}v_{i1}(k) + {K_{i2}v_{i2}(k)} +{}
    \\
     K_{i3}v_{i3}(k) + K_{i4}v_{i4}(k),
\end{multlined}
\end{equation}
where $K_{i1}$, $K_{i2}$, $K_{i3}$, and~$K_{i4}$ are non-negative constants depending on individual sheep agents, and the vectors $v_{i1}$, $v_{i2}$, $v_{i3}$, and $v_{i4}$ correspond to the four forces described above. 

{We assume that the $i$th sheep receives forces from all sheep in the circle with center $x_i(k)$ and radius $R > 0$. If there are no other sheep in this range, then we set $v_{i1}(k) = v_{i2}(k) = v_{i3}(k) = 0$. If we let $S_i(k)$ denote the set of the indices of the sheep within radius $R$ of the $i$th sheep at time~$k$, then the vectors $v_{i1}(k)$, $v_{i2}(k)$, $v_{i3}(k)$, and~$v_{i4}(k)$ are given by} 
\begin{align}
    v_{i1}(k) &= -\frac{1}{|S_i(k)|} \sum_{j \in S_i(k)} \frac{x_j(k) - x_i(k)}{\norm{x_j(k) - x_i(k)}^3}, \label{eq:sheep_v1}\\
    v_{i2}(k) &= \frac{1}{|S_i(k)|} \sum_{j \in S_i(k)} \frac{v_j(k-1)}{\norm{v_j(k-1)}}, \label{eq:sheep_v2}\\
    v_{i3}(k) &= \frac{1}{|S_i(k)|} \sum_{j \in S_i(k)} \frac{x_j(k) - x_i(k)}{\norm{x_j(k) - x_i(k)}} \label{eq:sheep_v3}\\
    v_{i4}(k) &= -\frac{x_d(k) - x_i(k)}{\norm{x_d(k) - x_i(k)}^3}.  \label{eq:sheep_v4}
\end{align}

\subsection{Farthest-Agent Targeting Method}\label{subsc:FAT}

The two methodologies for shepherding heterogeneous flocks are based on the shepherding algorithm for homogeneous flocks called the Farthest-Agent Targeting (FAT) method~\citep{tsunoda2019analysis}. In the FAT method, the shepherd guides the sheep farthest from a goal area. It is known that this simple strategy allows us to effectively guide the sheep agents. 

In order to describe the FAT method, we prepare some notations. Let the position of the shepherd at time~$k$ be denoted by $x_d(k)$. Then the update rule of the position~$x_d(k)$ within the FAT method is given by 
\begin{equation}
    x_d(k+1) = x_d(k) + v_d(k), 
\end{equation}
where $v_d(k)$ denotes the movement vector of the shepherd at time~$k$. The shepherd is assumed to receive the following three forces: an attractive force~$v_{d1}$ to the farthest sheep from the goal, a separation force~$v_{d2}$ to avoid colliding with nearest sheep to the shepherd, and a repulsion force~$v_{d3}$ to the goal area. Due to these three forces, the shepherd is able to get behind the sheep flock and guide it to the goal. Specifically, the movement vector~$v_d(k)$ is given by
\begin{equation}
    v_d(k) = K_{d1}v_{d1}(k) + K_{d2}v_{d2}(k) + K_{d3}v_{d3}(k),
\end{equation}
where $K_{d1}$, $K_{d2}$, and~$K_{d3}$ are non-negative constants\addd{, and the vectors $v_{d1}(k)$, $v_{d2}(k)$, and $v_{d3}(k)$ are given by}
\begin{align}
    \addd{v_1(k)} &\addd{= \frac{x_{t(k)}(k) - x_{\mathrm{d}}(k)}{||x_{t(k)}(k) - x_{\mathrm{d}}(k)||},} \label{eq:shepherd_v1}\\
    \addd{v_2(k)} &\addd{= -\frac{x_{n(k)}(k) - x_{\mathrm{d}}(k)}{||x_{n(k)}(k) - x_{\mathrm{d}}(k)||^3},} \label{eq:shepherd_v2}\\
    \addd{v_3(k)} &\addd{= -\frac{x_{\mathrm{g}}(k) - x_{\mathrm{d}}}{||x_{\mathrm{g}}(k) - x_{\mathrm{d}}||},} \label{eq:shepherd_v3}
\end{align}
\addd{where $t(k)$ denotes the index of the sheep farthest from the goal and $n(k)$ is the index of the sheep closest to the shepherd.}

\subsection{Shepherding Algorithm for Flock Containing Unre- sponsive Sheep}\label{subsc:method_himo}
The first guidance method for heterogeneous flocks that we present is the one proposed by~\cite{himo2022iterative}. Within their model, the sheep flock contains the following two types of sheep agents: responsive ones and unresponsive ones. While the responsive sheep agents are assumed to have positive gains $K_{i1}$, \dots, $K_{i4}$, the unresponsive ones are supposed to have zero gain for avoidance of the shepherd agent: i.e., $K_{i4}= 0$. 

Within the method by~\cite{himo2022iterative}, the shepherd is assumed to be given the position of all sheep as well as the type of each sheep (i.e., responsive or unresponsive). Based on these pieces of information, the shepherd is required to guide all the sheep agents to a prescribed goal area. The guiding behavior of the shepherd within the method can be divided into the following two types: collecting and guiding. First, the shepherd tries to guide responsive sheep around the unresponsive sheep.
Notice that unresponsive sheep are assumed to still have an attractive force to responsive sheep. Therefore, guiding all the unresponsive sheep around the responsive sheep allows the shepherd to indirectly guide the unresponsive sheep via guidance of responsive sheep. Therefore, after making sure that all the unresponsive sheep are close to responsive sheep, the shepherd can guide all sheep to the goal area. 
The guidance of a responsive sheep to unresponsive sheep within the method is performed by the FAT method described in Subsection~\ref{subsc:FAT}. Specifically, the shepherd guides a responsive sheep farthest from the unresponsive sheep to which the shepherd navigate the responsive sheep. 

It is shown in \cite{himo2022iterative} that the guidance methodology shows a high guidance success rate. Specifically, they define the shepherding success rate as the number of simulations of 100 different initial arrangements of
sheep that resulted in guiding the flock to the destination by a prescribed time limit. They have observed that their method outperforms the FAT method  regardless of the number of unresponsive sheep for any of the scenarios of $N = 10$,
$N=30$, and $N=50$, where $N$ denotes the total number of sheep. 

\subsection{Shepherding Algorithm for Flock Containing Variant Sheep}\label{subsc:method_fuji}

In this subsection, we describe the heterogeneous flock and guidance method proposed by~\cite{fujioka2022shepherding}. Let us first describe their model. 
Within their model, they introduce a sheep agent called a variant sheep, which makes the flock heterogeneous. Within their model, the variant sheep is assumed not to receive more than one force among the separation, alignment, attraction, and avoidance. For this reason, there are a total of $14$ different kinds of variant sheep: those receiving only one force, two forces, and three forces, respectively. They further assume that one and only one type of variant sheep can exist. 
To model the variant sheep, they simply set the gain of the forces that the variant sheep is not receiving to be zero, which is similar to the model by~\cite{himo2022iterative} described in  Subsection~\ref{subsc:method_himo}. 

\begin{figure}[tb]
    \centering
    \includegraphics[width=.85\linewidth]{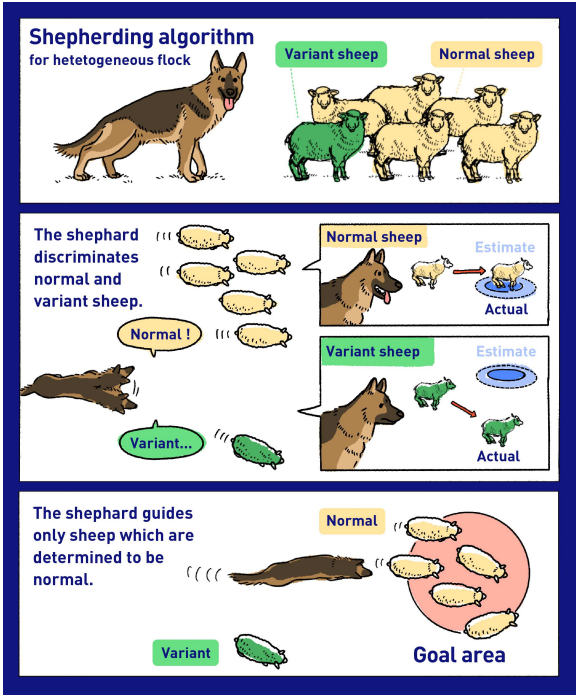}
    \caption{The outline of the method proposed by~\cite{fujioka2022shepherding}. Illustration by Chihiro Kinoshita.}
    \label{fig:outline}
\end{figure}

In this method, the shepherd is assumed to be able to  periodically obtain the positions of all sheep. It is also assumed that a set of estimated coefficients of the normal sheep (i.e., the sheep that are not variant) is given to the shepherd. It should be remarked that the estimate is not necessarily correct.
Based on the estimate, the shepherd then predicts  the trajectories of the sheep. In periodic acquisition of sheep positions, the shepherd measure the distance between the predicted trajectory and the actual, measured position. If the distance is larger than a threshold value, the sheep is determined to be variant by the shepherd. Then the shepherd guide only sheep which are determined to be normal. We show the outline of the method in Fig.~\ref{fig:outline}.

In \cite{fujioka2022shepherding}, the authors have compared the guidance performance of their method and the FAT method. They have confirmed that, due to the incorporated discrimination mechanism, their guidance method generally performs better than the FAT method. Specifically, although the FAT method tends to show smaller guidance success rates as the number of the variant sheep in the flock increases, their method exhibited relatively high performance (guidance success rates $>$ 60\%) irrespective of the type of variant sheep. 

\section{PROSPECTS}\label{sc:prospects}


In this section, we describe the prospects of shepherding research for heterogeneous flocks.

As we described in Section~\ref{sc:methods}, some guidance methods have been proposed so far. \cite{himo2022iterative} proposed the shepherding algorithm for a flock containing unresponsive sheep. Our research group extended this work by defining $14$ types of variant sheep and proposed a method that can be applied to all types.
The development of such a method for guiding a heterogeneous flock will expand the potential applications. However, there are still some problems with the methods as described below. 

\subsection{Generalization of Sheep Model}\label{subsc:sheep_expansion}
First, let us consider the generalization of the sheep model. The first possible extension of the sheep model is to further increase the variety of variant types by changing the gains. For example, if the coefficient of repulsion $K_{i4}$ is set to any value greater than $0$, the magnitude of the response to the shepherd can be varied. If set to a negative number, it is possible to define variant sheep that attract to the shepherd. This kind of induction on variant sheep allows us to develop a method that is robust against sheep that move unexpectedly.

Another option is to increase the number of variant types present at the same time. In our method, the shepherd calculates the threshold value based on the distance between the actual and predicted positions of all sheep, but this threshold may not be set appropriately when multiple types of variant sheep are present at the same time. Therefore, one of our future works is to develop a guidance method for such a situation.

Also, we can consider a more significant change to the sheep model. In this paper, we presented sheep model based on Boid model. However, we would like to redefine a heterogeneous flock using a model other than the Boid model. For example, \cite{go2021solving} assumed that the sheep receives the following five forces: ``previous force'' to attract to previous direction, ``attraction force'' to the center of mass of the group, ``repulsion force'' to avoid other sheep, ``repulsion force'' to avoid the shepherd, and ``random error force''. This change in the sheep model may change the method of efficient flock guidance. Another future task is to develop a method that can correctly guide flocks for any sheep model.

\subsection{Expansion of the Shepherd Model}
\begin{figure}[tb]
  \centering
  \subfloat[One shepherd]{\includegraphics[width=.6\linewidth]{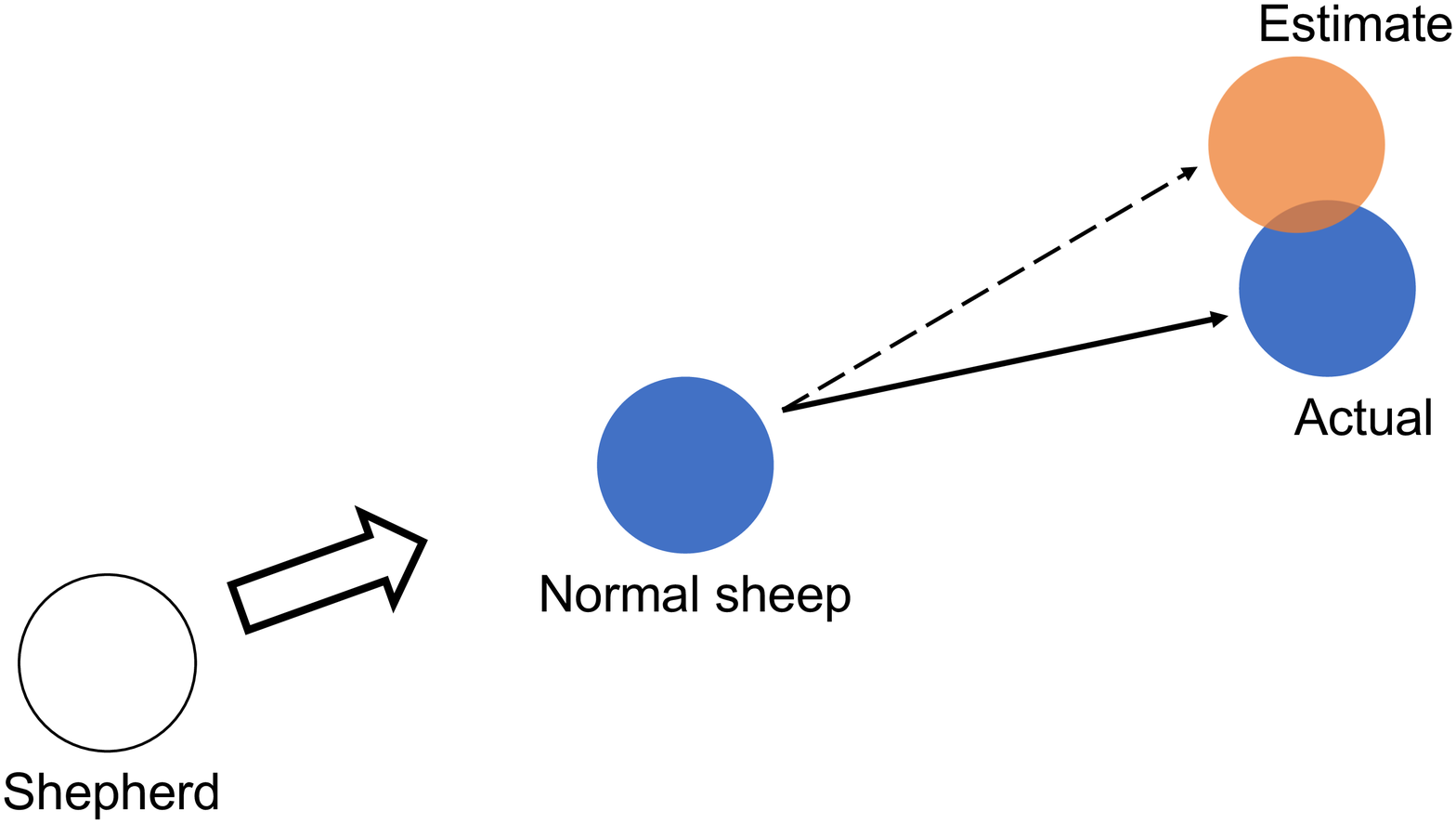}
  \label{fig:one_shepherd}}
  \\
  \subfloat[Two shepherds]{\includegraphics[width=.6\linewidth]{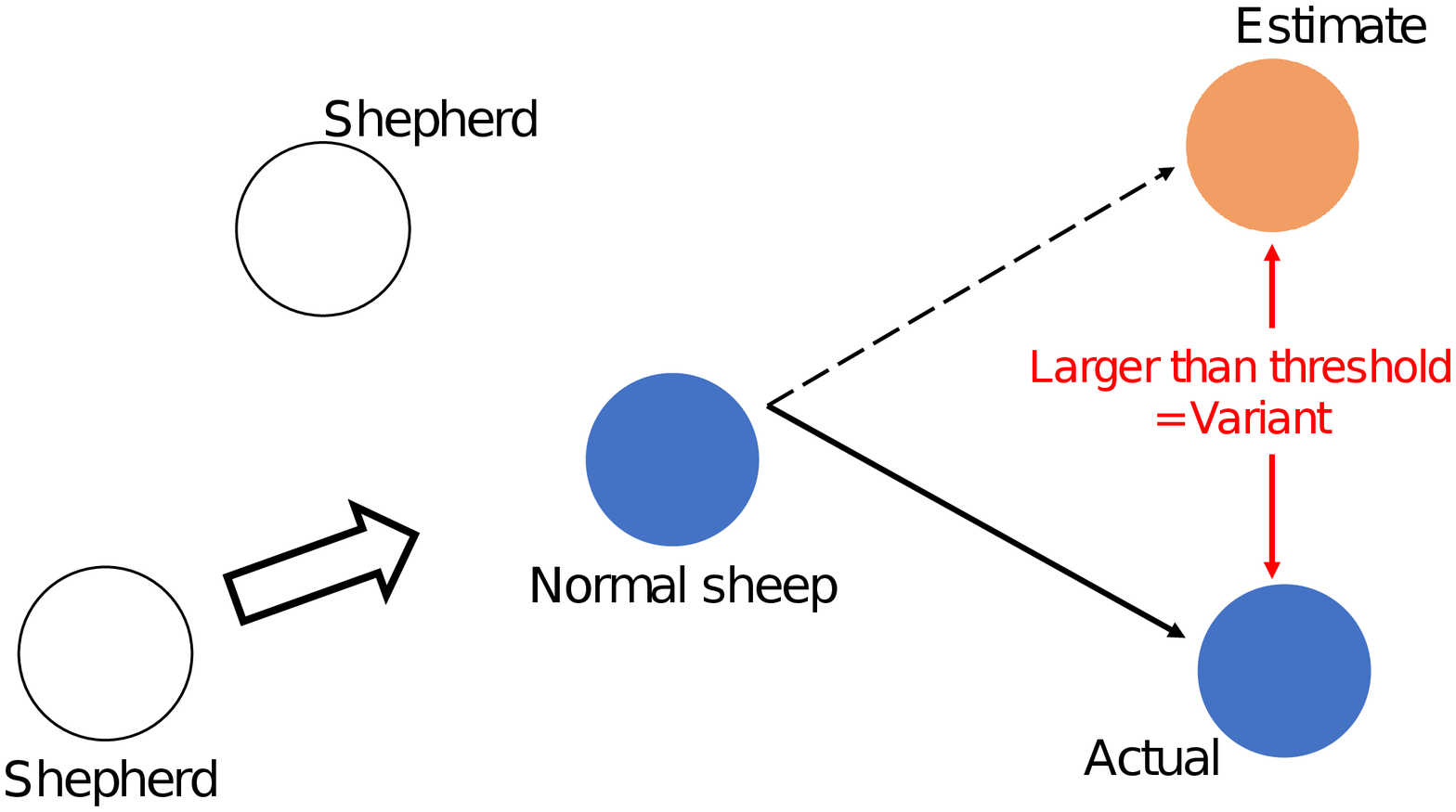}
  \label{fig:two_shepherds}}
  \caption{If there is one shepherd, the shepherd can estimate trajectories of sheep properly. However, if there are more than one shepherd, it is difficult to consider the impact of other shepherds on the sheep's trajectory.
}
  \label{fig:shepherd}
\end{figure}

Next, let us discuss the shepherd model and the movement laws. It is unclear whether our determine method is effective when there are several shepherds. If the shepherds cannot communicate with each other, the trajectory predicted by one shepherd may be inaccurate due to the influence of the other shepherds (See Fig.~\ref{fig:shepherd}).
If shepherds can communicate with each other (e.g., if they can obtain the positions of other shepherds), such inaccuracies in predicted trajectories are unlikely to occur. However, when the FAT method is used, there is a high probability that all shepherds will chase the same sheep. To prevent this, it would be necessary to use a different movement law for multiple shepherds~\citep{lien2005shepherding}. We would also like to evaluate the effectiveness of our determine method when used in these methods.

{\subsection{Theoretical Analysis}}
{In all of the methods described in Section~\ref{sc:methods}, the effectiveness of the method is measured by simulating the success rate of guidance using multiple initial positions. The agent models are nonlinear and, therefore, it is difficult to prove that the guidance is successful. However, we believe that the search for a proof method will allow us to investigate the exact effectiveness of the method.}

{Let us discuss how the proof can be done. The cause of the difficulty of the proof is the non-linearity of the agent's model. One idea is to simplify the model. For example, in the FAT method described in Subsection~\ref{subsc:FAT}, we can assume that the shepherd always exists at a fixed distance from the farthest sheep, or we can consider only minimal interactions by setting the number of sheep to be equal to two.}
{Such a proof would allow us to analyze Degree of heterogeneity for successful guidance. For example, when the coefficient of a variant sheep is $\alpha K_i$ using an arbitrary real number $\alpha$ for the coefficient $K_i$ of a normal sheep, it will be possible to identify the range of $\alpha$ with high guidance succeeds. Further consideration of these results will enable us to develop a method with a higher guidance success rate.}

\section{CONCLUSION}\label{sc:conclusion}
In this paper, we have described conventional researches and have presented future perspectives on a heterogeneous flock and its guidance methods. The definition of heterogeneous flocks and the development of movement laws of the shepherd are still far from complete. However, we believe that further research on heterogeneous flocks will lead to a better understanding of flock control. Through further research, we hope to expand the applicability of flock control.


                                                   










\end{document}